\documentclass[usenatbib]{mn2e}

\usepackage[utf8x]{inputenc}
\usepackage{bm,latexsym,tabularx,dcolumn}
\usepackage{amsmath,amsfonts,amssymb}
\usepackage{graphicx,epsfig,color,subfigure,sidecap,float}
\usepackage{longtable}
\usepackage{natbib}

\newcommand{\bea}{\begin{eqnarray}}
\newcommand{\eea}{\end{eqnarray}}
\newcommand{\be}{\begin{equation}}
\newcommand{\ee}{\end{equation}}
\newcommand{\f}{\frac}
\newcommand{\df}{\dfrac}
\newcommand{\dl}{\delta}

\newcommand{\bc}{\begin{center}}
\newcommand{\ec}{\end{center}}
\newcommand{\mcn}{\multicolumn}

\newcommand{\T}{\rule{0pt}{3.6ex}}

\newcommand{\B}{\rule[-1.0ex]{0pt}{0pt}}

\title{Understanding the non-linear clustering of high redshift galaxies}
\begin{document}
\author[C. Jose et al.]
{Charles Jose$^{1,2}$\thanks{charles.jose@.durham.ac.uk}, 
Carlton M. Baugh$^{1}$, 
Cedric G. Lacey$^{1}$ and \newauthor
Kandaswamy Subramanian$^3$\\
$^1$Institute for Computational Cosmology, Department of Physics, University of Durham, South Road, Durham DH1 3LE, UK \\
$^2$Department of Physics, SB College, Changanassery, Kottayam 686101, India \\
$^3$IUCAA,Post Bag 4, Pune University Campus, Ganeshkhind, Pune 411007, India
}

\pubyear{2017}

\maketitle


\maketitle

\begin{abstract}
We incorporate the non-linear clustering of dark matter 
halos, as modelled by \citet{jose_16} into the halo model to better 
understand the clustering of Lyman break galaxies (LBGs) 
in the redshift range $z=3-5$.
We find that, with this change, the predicted LBG clustering 
increases significantly on 
quasi-linear scales ($0.1 \leq r\,/\,h^{-1} \,{\rm Mpc} \leq 10$) compared 
to that in the linear halo bias model. 
This in turn results in an increase in the clustering of LBGs by an order of 
magnitude on angular scales $5'' \leq \theta \leq 100''$.  
Remarkably, the predictions of our new model on the whole remove the 
systematic discrepancy between the linear halo bias predictions 
and the observations. 
The correlation length and large scale galaxy bias of LBGs 
are found to be significantly higher in the non-linear halo bias 
model than in the linear halo bias model. 
The resulting two-point correlation function retains an  
approximate power-law form in contrast with that computed using 
the linear halo bias theory.
We also find that the non-linear clustering of LBGs increases with 
increasing luminosity and redshift.
Our work emphasizes the importance of using non-linear halo bias  
in order to model the clustering of high-z galaxies to probe the physics of galaxy 
formation and extract cosmological parameters reliably.
 
\end{abstract}


\section{Introduction}

The halo model of large scale structure is a successful formalism 
for predicting and interpreting the clustering of 
dark matter halos and the galaxies associated with them \citep{cooray_sheth_02}. 
In the halo model the galaxy correlation function ($\xi_{\mathrm g}(r)$) is the sum of 
two terms, the 1-halo term and the 2-halo term. 
The 1-halo term accounts for the contributions to the clustering from 
galaxy pairs residing within the same dark matter halo and dominates 
the galaxy clustering on small scales. 
Since visible galaxies are formed in dark matter halos and 
subhalos, the shape of the 1-halo term is determined by dark matter 
halo substructure \citep{berlind_02}. 
The 2-halo term, that accounts for the correlation 
between galaxies residing in distinct halos, dominates the clustering 
of galaxies on scales larger than their typical halo virial radius. 
The crucial component of the 2-halo term is the halo bias that 
determines how dark matter halos trace the dark matter on large 
scales \citep{kaiser_1984,cole_kaiser_1989, bond_91}.

Interestingly, the predicted galaxy correlation functions at low 
redshifts have an almost power-law form ($\xi_{\mathrm g}(r) = (r/r_0)^{-\gamma}$), 
with a subtle feature on scales ($r \sim 1-2  h^{-1} \text{Mpc}$) corresponding 
to the transition from the 1-halo to the 2-halo term 
\citep{zehavi_weinberg_04, berlind_02,kravtsov_04, watson_berlind_11}. 
The halo model has been highly successful in 
explaining many low redshift clustering measurements over a wide 
range of galaxy-galaxy separations \citep{zehavi_weinberg_04,
zehavi_zheng_05,conroy_wechsler_06, zheng_zehavi_09, zehavi_zheng_11,
parejko_sdss_2013,guo_2013_sdss}.
The clustering measurements for high redshift ($3\leq z \leq 5$) galaxy samples  
shows stronger departures from a power law, particularly around the 
transition from the 1-halo to 2-halo terms \citep{ouchi_hamana_05_acf,
lee_giavalisco_06_acf,hildebrandt_09_acf,cooke_2013_acf}
Nevertheless, the measured galaxy correlation functions in this redshift 
range are typically approximated as a power-law \citep{kashikawa_06_acf,
savoy_11_acf, bian_2013_acf, durkalec_2015_acf,park_2016, 
harikane_2016_acf,ishikawa_2016}. Interestingly, the measured clustering of 
galaxies retains an approximate power-law form even at a higher 
redshift $ \sim 7$ \citep{barone_2014_acf}. 

However, models of the clustering of very high redshift ($z \geq 3$) galaxies 
using the standard halo model predict a much stronger 
deviation from a power law  
than is suggested by the observations 
\citep{ouchi_hamana_05_acf,lee_09, charles_13_clustering}. 
In particular, \cite{charles_13_clustering} (hereafter J13) 
point out that the predicted angular clustering of 
Lyman Break Galaxies (LBGs) on angular scales $10'' \leq \theta \leq 100''$ 
is lower than the observed clustering by an order of magnitude. 
These angular scales (corresponding to comoving length of $0.5-10$ Mpc) 
are bigger than the virial radii of the typical dark matter halos 
at that redshift, but smaller than scales where the linear theory 
is valid, and therefore referred to as quasi-linear scales. 
On these scales, the significant contribution to the galaxy clustering comes 
from the 2-halo term. J13 modelled the 2-halo term in this regime using the 
large scale linear halo bias \citep{mo_white_96,sheth_tormen_99,tinker_10}. 
At $z=0$, deviations from the linear halo bias approximation on quasi-linear 
scales are found to be of the order of only a few percent 
\citep{tinker_weinberg_05,bosch_13}. However, such a weak scale dependence 
of the halo bias is not sufficient to explain the 
discrepant clustering strength of high-z LBGs. To summarize, the halo model 
predictions of high-z LBG clustering which use linear halo bias 
show substantial departures from a power-law and underpredict the 
observed correlation functions on quasi-linear scales (J13).

Here we investigate whether considering the scale-dependent non-linear bias 
of high-z dark matter halos on quasi-linear scales removes the tension 
between the halo model predictions and the observed LBG clustering. 
For this, we use the model introduced by \cite{jose_16} (hereafter J16) 
who showed that the linear halo bias approximation breaks 
down on quasi-linear scales at high-z. In particular, they showed that 
the scale-dependence of non-linear halo bias in this region is much stronger 
at high-z compared to low-z.

There are previous works, especially at 
low redshifts ($z \sim 0$), which show the importance of using 
a weakly scale-dependent non-linear halo bias on quasi-linear scales for a 
better comparison of theoretical predictions with observed 
data \citep{cooray_04,zehavi_weinberg_04,bosch_13}. 
The consequences of the non-linear clustering of mini-halos 
on quasi-linear scales during the cosmic dark ages ($z \geq 6$) and the  
implications for future observations of this epoch are 
discussed by \citealt{iliev_scannapieco_03} and \citealt{reed_09}. 
Here, we focus on the clustering of galaxies in the redshift range 
$3-5$, thereby bridging the gap between these two distinct epochs in the 
cosmic history. 
This is the first work, to our knowledge, that uses the non-linear bias 
of dark matter halos on quasi-linear scales to better understand the measured 
clustering of LBGs at $3 \leq z \leq  5$.

The organization of this paper is as follows. In the next section 
we present our model for the halo 
occupation distribution and the non-linear clustering of high-z dark matter halos. 
In Section 3, we compute the two point spatial and angular correlation functions 
of LBGs and compare them with observations. We present our conclusions in the 
final section. For all calculations we adopt a flat $\Lambda$CDM 
universe with cosmological parameters consistent with the 9 year Wilkinson 
Microwave Anisotropy Probe observations \citep{hinshaw_2011_wmap9}.
Accordingly we assume a Hubble parameter $h= 0.70$, a baryon density $\Omega_b=0.0463$, cold dark matter density $\Omega_c= 0.233$, density of massive neutrinos 
$\Omega_\nu = 0.0$, fluctuation normalization $\sigma_8=0.821$ and spectral index $n_s= 0.972$ (where $\Omega_i$ is the background density of any species `i' in units of critical density $\rho_{c}$). 

\section{The clustering of high redshift LBGs}
Here we discuss the two key components of our galaxy clustering 
model: (i) the halo occupation distribution and (ii) the model for the non-linear 
clustering of dark matter halos.  
\subsection{The halo occupation distribution}
The halo occupation distribution (HOD) is an important ingredient for 
computing the two point correlation function of any galaxy sample. 
The HOD gives mean number of galaxies of a given type in a dark 
matter halo of mass $M$ and is usually presented in a parameterized 
form \citep{seljak_00_HOD,bullock_wechsler_02_HOD,kravtsov_04,
hamana_04,zehavi_zheng_05,hamana_06,conroy_wechsler_06}. 
Galaxy clustering can be calculated by combining the HOD with 
dark matter clustering, dark matter halo 
abundance, halo bias and the halo density profile. 

In order to compute the clustering of LBGs selected to be brighter than some luminosity 
threshold, we use the HOD computed using the galaxy formation model 
of J13 for illustrative purposes (see also \citealt{samui_07}). 
In this model, each dark matter halo can host a central galaxy and satellite
galaxies. 
The central galaxy is put at the center of the halo and 
satellite galaxies are distributed around the central galaxy 
following the dark matter density profile. 
The satellite galaxy occupation, which is the mean number of satellite 
galaxies in a dark matter halo, is computed using the conditional or 
progenitor mass function \citep{lacey_cole_93, cooray_sheth_02}. 
J13 then incorporate a physically motivated model for the total star 
formation rate in a dark matter halo to compute the UV luminosity 
of the LBGs hosted by the halo. 
The parameters of the model are constrained by fitting the observed 
UV luminosity functions of LBGs from \citet{bouwens_07_LF_z46},
\citet{reddy_08_LF} and \citet{bouwens_12_LF}. 
Thus J13 constrains the relation between the UV luminosity and halo mass 
of high-z LBGs from which the HOD of LBGs brighter than a given 
luminosity is computed. 

The mean number of LBGs brighter than apparent AB magnitude 
$ m$ inside a dark matter halo of mass $M$, separated into central 
and satellite components, can be written as  
\be 
N_{\text{g}}(M, m,z) = N_{\text{cen}}(M, m,z)+N_{\text{s}}(M, m,z). 
\label{eq:totgal}
\ee
Here, $N_{\text{cen}}(M, m,z)$ and $N_{\text{s}}(M, m,z)$ are respectively the 
average number of central and satellite galaxies in a halo of 
mass $M$, satisfying the luminosity threshold condition. 
The separation of the HOD of LBGs into central and satellite 
components is crucial for computing the clustering  
\citep{kravtsov_04, zheng_05,cooray_ouchi_06,conroy_wechsler_06}.  
Further details about the computation of the central and satellite 
contribution to the HOD can be found in J13. 

\begin{figure}
\includegraphics[trim=0cm 0cm 0cm 0cm, clip=true, width =8cm, height=16cm, angle=0]
{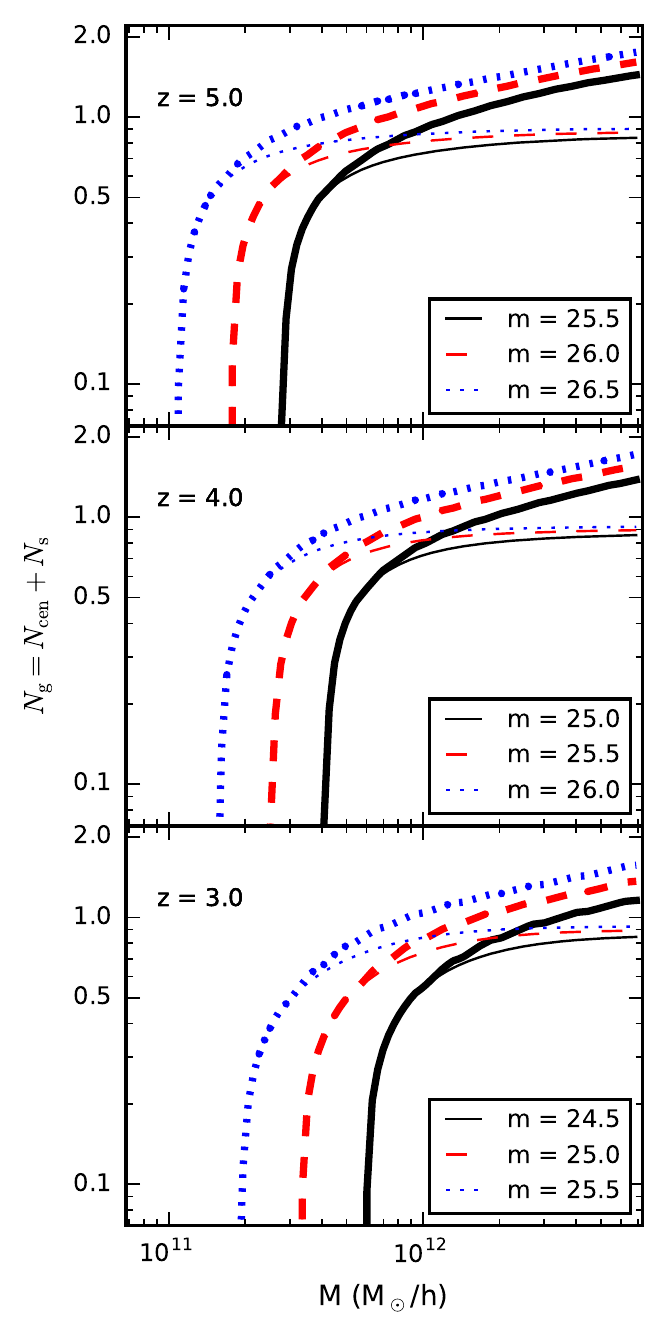} 
\caption[]
{
The halo occupation distribution, $ N_{\text{g}}(M,m,z) $, as a function 
of the mass of the host halo, given by J13 at various redshifts and 
limiting magnitudes ($m$) as labelled (same as Fig.~7 of J13). 
In each panel the thin curves correspond to $N_{\text{cen}}(M,m,z)$. 
The thick curves give the total occupation 
$N_{\text{g}}(M,m,z) = N_{\text{cen}}(M,m,z)+N_{\text{s}}(M,m,z)$. The mean occupation of 
the central galaxies is always less than unity. For a fuller 
description see J13.  
}
\label{fig:Ng}
\end{figure}

In Fig.~\ref{fig:Ng} (same as Fig.~7 in J13) the average occupation number 
of LBGs is plotted as function of the mass of the parent halo computed using 
the model of J13 for $z =3-5$ and for three 
apparent magnitude thresholds, for which clustering measurements are 
available \citep{hildebrandt_09_acf}. 
Also shown is $N_{\text{cen}}(M, m,z)$, the 
mean occupation number of central LBGs.  
We note that for all redshifts and threshold magnitudes shown the mean 
occupation number of central LBGs plateaus at a value less than unity. 
Furthermore, the average halo mass of LBGs in any sample varies from 
$3 \times 10^{11} - 1.5 \times 10^{12} M_\odot$ (see Table~\ref{tab1} and also J13). 
These galactic dark matter halos at high-z correspond to $2\sigma - 4\sigma$ 
fluctuations, whereas at $z=0$ halos of comparable mass collapse 
from perturbations that are less than 1 $\sigma$. 
Thus galactic mass dark matter halos at $z=4$ are 
much `rarer' than those at $z=0$.
 
\subsection{Analytic model for the non-linear clustering of high-z halos}

Now we briefly describe our model for the non-linear clustering of 
high-z dark matter halos on quasi-linear scales. 
The dark matter halos are assumed to be spherical 
regions with an over density $\Delta= 200$ times 
the background density of the Universe. Then, the virial radius $r_{200}$ of 
any halo of mass M is given by $M = (4/3) \pi r^3_{200} \rho_c \Delta$. 
In the linear bias approximation, the cross-correlation between 
halos of mass $M'$ and $M''$ is given by \citep{cooray_sheth_02} 
\be
\xi_{hh}(r|M', M'',z) = b(M',z) b(M'',z) \xi_{\mathrm{mm}}(r,z), 
\label{eq:lin_bias_approx}
\ee
where $\xi_{\mathrm{mm}}(r,z)$ is the two point correlation function of the matter 
density field at redshift $z$ and $b(M,z)$ is the scale-independent linear bias 
of halos of mass $M$. Eqn.~\ref{eq:lin_bias_approx} is valid on large 
scales, where density fluctuations still grow linearly in redshift. 

On quasi-linear scales and for rarer halos \cite{jose_16} 
improved the model described in Eq.~(\ref{eq:lin_bias_approx}) by 
replacing the scale-independent linear halo bias with a scale-dependent 
non-linear halo bias given by
\be
b_{\mathrm{nl}}(r,M,z) =  b(M,z) \zeta(r,M,z).
\label{eq:nlbias}
\ee
Here $\zeta(r,M,z)$ is the scale-dependent part of the 
non-linear halo bias while $b(M,z)$ is the linear halo bias measured 
on large scales. 
The functions $b(M,z)$ and $\zeta(r,M,z)$ 
are calibrated by fitting to N-body simulations. 
We also note that, by definition, $\zeta(r,M,z) \longrightarrow 1$ 
on very large scales.

The expression for the large scale halo bias $b(M,z)$ is given by the 
fitting function of \citet{tinker_10}  
\be
b(M,z) = b(\nu(M,z)) = 1 - A \df{\nu^a}{\nu^a + \delta_c^a} + B \nu^b + C \nu^c,
\label{eq:tbias}
\ee 
where $\nu(M,z) = \dl_c/\sigma(M,z)$ is the peak-height, $ \dl_c=1.686$ is 
the critical density for spherical collapse 
and $\sigma(M,z)$ is the linear theory variance of matter fluctuations on a 
mass scale $M$ at redshift $z$.  
\cite{jose_16} found that the \cite{tinker_10} fitting function slightly 
overpredicts the large scale bias of the rarest dark matter halos. 
This is probably due to the different algorithms used in these 
studies to identify dark matter halos.  
Therefore, \cite{jose_16} refitted Eq.~(\ref{eq:tbias}) finding the best fitting  
parameters $A=1.0, a= 0.0906, B=-4.5002, b= 2.1419, 
C= 4.9148$ and $c= 2.1419$. 
Here we will use the parameter values given by \cite{jose_16}.

The scale-dependent function $\zeta(r,M,z)$ depends on $r$, $M$ and $z$ through 
four quantities: $\nu(M,z)$, $\xi_{\mathrm{mm}}(r,z)$, an effective power-law index of $\sigma(M)$, $\alpha_m(z)$, and the matter density of the universe 
at a given redshift, $\Omega_m(z)$.
This effective power-law index, $\alpha_m(z)$, is defined as
\be
\alpha_m(z) = \df{\log(\dl_c)}{\log[M_{\text{nl}}(z)/M_{\text{col}}(z)]}. 
\label{eqn:alpha}
\ee
Here $M_{\text{nl}}$ is the non-linear mass scale where the peak-height 
$\nu(M,z) = \dl_c$ 
($\sigma(M,z) = 1$) and $M_{\text{col}}$ is the collapse mass scale where  
$\nu(M,z) = 1$ ($\sigma(M,z) = \dl_c$). These masses can be computed numerically 
for any given cosmological model to evaluate $\alpha_m(z)$. 
The matter density of the universe at any redshift can be written as 
\be
\Omega_m(z) = \df{\Omega_m(0) (1+z)^3}{\Omega_m(0) (1+z)^3 + \Omega_\Lambda(0)}, 
\ee
where $\Omega_m(0)$ and $\Omega_\Lambda(0)$ are the densities of matter and dark energy 
at $z=0$ in units of the critical density. 

The form for $\zeta(r,M,z)$ expressed in terms of these quantities was 
found by J16 to be well fitted by
  
\begin{align}
& \zeta(\xi^{\rm sim}_{\rm mm},\nu, \alpha_m,\Omega_m(z)) =  \nonumber \\ 
&~~~~\left(1 + K_0 \log_{10}\left(1+{\xi_{\rm mm}}^{k_1}\right) \nu^{k_2} (1+k_3/\alpha_m) {\Omega_m(z)}^{k_4} \right) \times \nonumber\\
&~~~~\left(1 + L_0 \log_{10}\left(1+{\xi_{\rm mm}}^{l_1}\right) \nu^{l_2} (1+l_3/\alpha_m) {\Omega_m(z)}^{l_4} \right),  
\label{eq:b_r_full} 
\end{align}
with $K_0 = 0.1699, k_1 = 1.194, k_2 = 4.311, k_3 =−0.0348, k_4 = 17.8283, L_0 = 2.9138, 
l_1 = 1.3502, l_2 = 1.9733, l_3 =-0.1029$ and $l_4 = 3.1731$. 

\begin{figure}
\includegraphics[trim=0cm 0.0cm 0cm 0.0cm, clip=true, width =8.0cm, height=13.0cm, angle=0]
{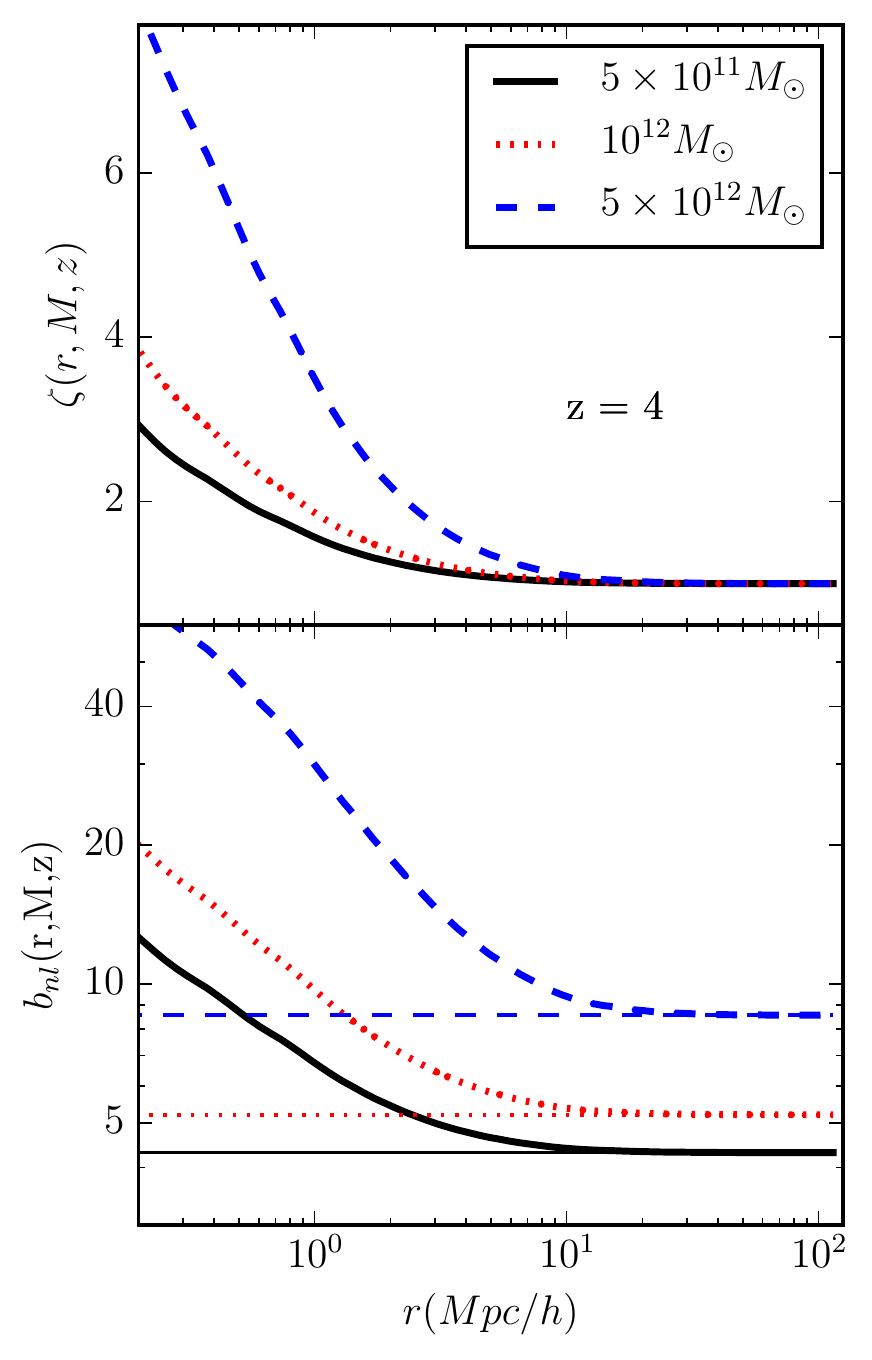} 
\caption[]
{Top panel: the scale dependence $\zeta(r,M,z)$ of the non-linear halo bias 
as a function of the halo separation for three halo masses 
at $z=4$. Bottom panel: the non-linear halo bias (thick lines)
$b_{\mathrm{nl}}(r,M,z) = b(M,z) \zeta(r,M,z)$ at the same 
redshift. The thin horizontal lines are the scale-independent linear 
halo bias $b(M,z)$ for the same masses. }
\label{fig:bnl}
\end{figure}

In the top panel of Fig.~\ref{fig:bnl}, we plot $\zeta(M, r,z)$ for  
halos as a function of separation for three different halo masses that 
host LBGs at $z=4$. The bottom panel shows the corresponding non-linear halo bias 
$b_{\mathrm{nl}}(r, M,z)$ of the halos along with the scale-independent 
linear halo bias (thin horizontal line) given by Eq~(\ref{eq:tbias}).
It is clear from Fig. \ref{fig:bnl} that $\zeta(M, r,z)$ increases 
significantly from unity on small scales. 
As a result, the non-linear halo bias is not constant and is strongly scale-dependent 
on quasi-linear scales ($0.5-10$ $h^{-1}$ Mpc). 
One can see that there is a significant boost in the halo bias at high-z, even on  
scales of $5-10$ $h^{-1}$ Mpc. For example, at $z=4$ the bias of 
dark matter halos of mass $M = 5 \times 10^{12} M_\odot$ is increased 
by a factor $4.5$ at a halo separation $r= 1$ $h^{-1}$ Mpc, compared 
to the linear bias of those halos. This will boost the clustering strength 
of these halos by a factor 20. 
Furthermore, $\zeta(M, r,z)$ increases with halo mass suggesting 
that the non-linear bias is stronger for rarer halos. 
As we shall see later, such a strong scale-dependent non-linear 
halo bias results in a remarkable change in the predicted shape of the 
LBG correlation function compared to the predictions using linear halo bias.

The halo-halo correlation function is then obtained by replacing the linear halo 
bias with the non-linear halo bias, $b_{\mathrm{nl}}(r,M,z)$; 
\be
\begin{split}
1+\xi_{\mathrm{hh}}(r|M',M'', z) =& [1+ b_{\mathrm{nl}}(r,M',z) b_{\mathrm{nl}}(r,M'',z)  \\
              & \xi_{\mathrm{mm}}(r) ]~ \Theta[r-r_{\mathrm{min}}(M',M'')]
\end{split}
\label{eq:xi_hh_full}
\ee
Here the $\Theta$ function incorporates halo exclusion in our calculations. 
The halo exclusion  ensures that the correlation between any pair 
of halos of masses $M'$ and $M''$ goes to $-1$ if $r <r_{\mathrm{min}}(M',M'')$ 
where $r_{\mathrm{min}} = \text{max}[r_{200}(M'),r_{200}(M'')]$ \citep{bosch_13}. 
Here $r_{200}(M) = \left(3M/4 \pi \rho_c \Delta \right)^{1/3}$ is the virial 
radius of a dark matter halo of mass $M$ with $\Delta = 200$. 

\begin{figure*}
\includegraphics[trim=0cm 0.0cm 0cm 0.0cm, clip=true, width =15.0cm, height=5.0cm, angle=0]
{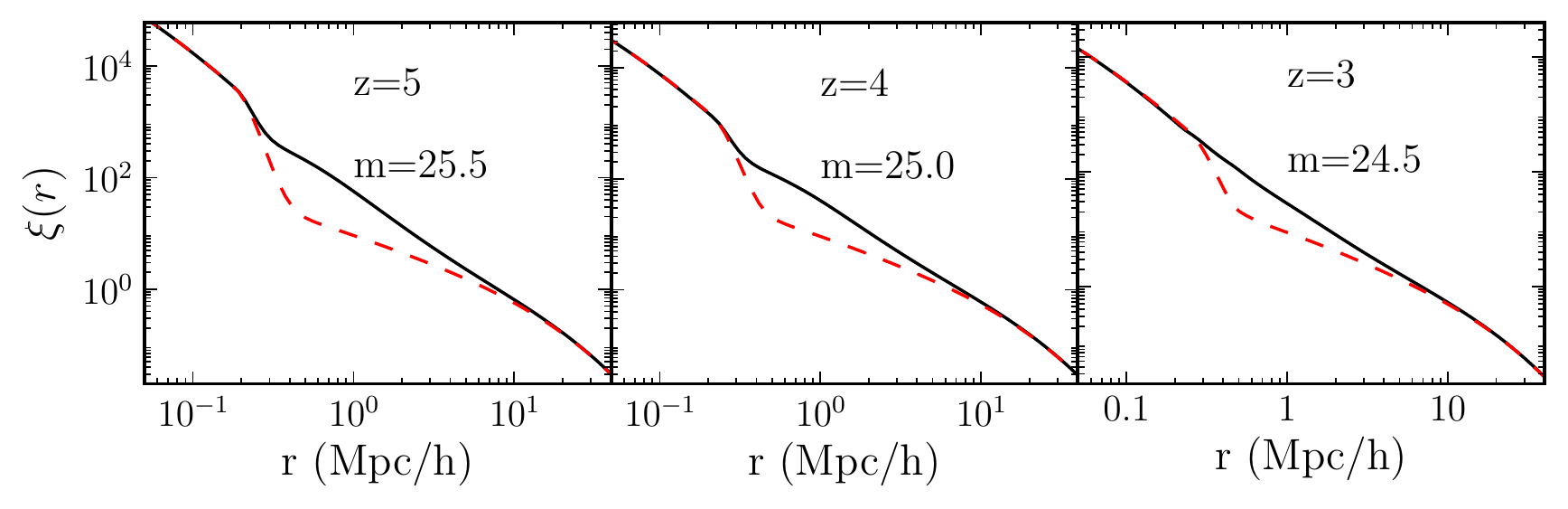} 
\caption[]
{
The spatial correlation functions, $\xi_{\mathrm g}$, of high-z LBGs at redshifts 3, 4 and 5 
and for a given apparent magnitude limit as labelled.  
The red dashed and black solid curves are respectively the predictions of scale-independent 
linear halo bias and scale-dependent non-linear halo bias models.  
}
\label{fig:xir}
\end{figure*}
\section{The correlation functions of LBGs} 

We now incorporate the non-linear halo bias into the halo model of 
large scale structure to compute the correlation functions of LBGs. 
As discussed above, for this we use the HOD (shown in Fig.~\ref{fig:Ng}) of 
LBGs computed using the model of J13. 
The galaxy correlation functions  
have a contribution from the 2-halo term that describe the clustering 
between galaxies hosted in distinct halos and a 1-halo term that accounts for 
the clustering from galaxies residing in the same dark matter halo. 
The non-linear halo bias modifies the 2-halo term by boosting the clustering of 
galaxies on quasi-linear scales. 
The 1-halo term, however, remains unchanged because the non-linear halo bias 
does not alter the distribution of galaxies inside a dark matter halo. 

\begin{figure*}
\includegraphics[trim=0cm 0.0cm 0cm 0.0cm, clip=true, width =15.0cm, height=15.0cm, angle=0]
{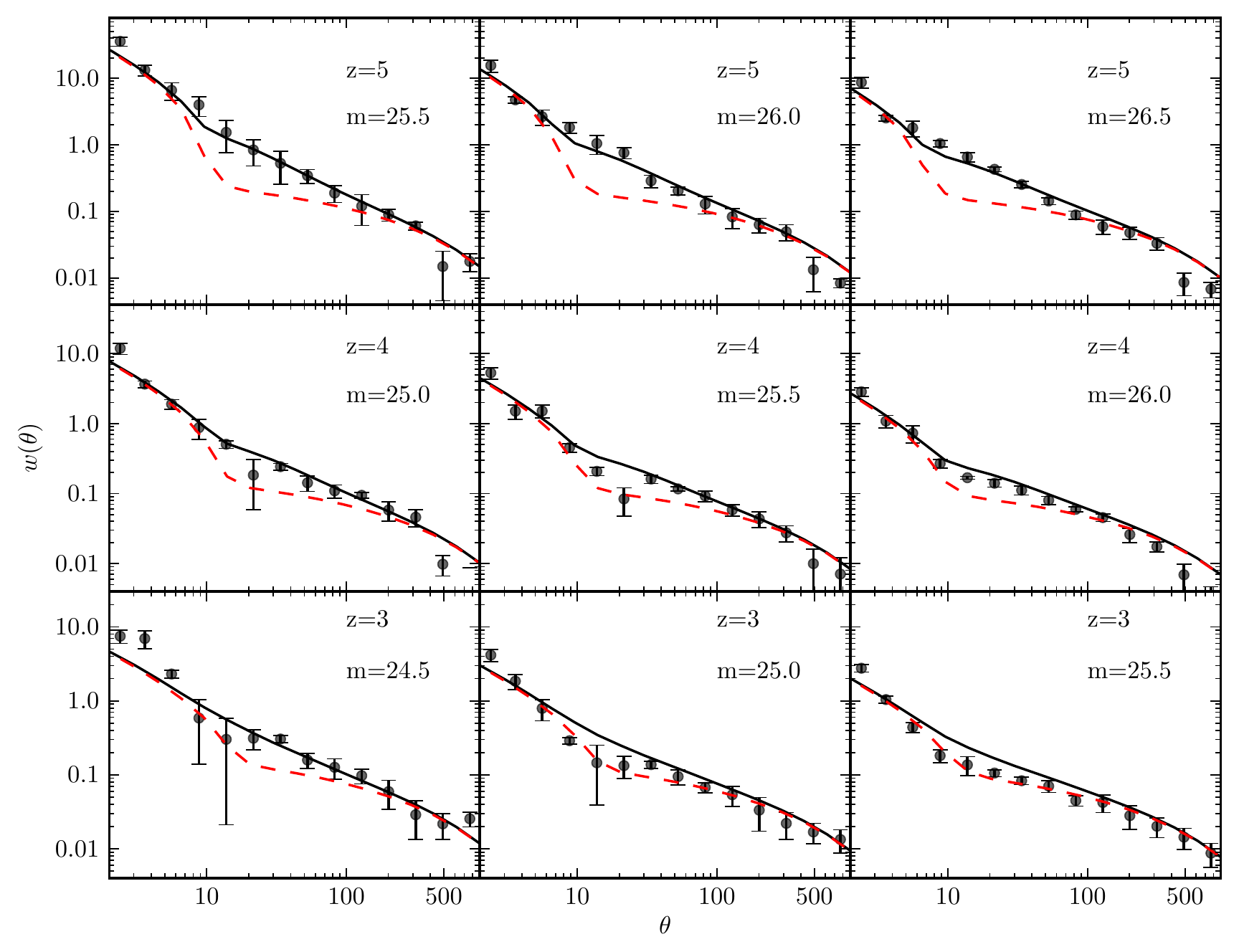} 
\caption[]
{
The angular correlation functions of LBGs predicted using non-linear (solid black) 
and linear (dashed red) halo bias models at redshifts 3, 4 and 5 and for various limiting 
magnitudes as labelled. 
In each row there are three panels showing the clustering 
predictions of LBGs samples with three apparent magnitude limits. The observational 
clustering measurements are taken from \protect \citet{hildebrandt_09_acf}  
}
\label{fig:acf}
\end{figure*}

\subsection{The 2-halo term}
When computing the 2-halo term with non-linear halo bias, 
we assume that the halo density profile is 
sufficiently peaked that it does not affect the 2-halo term on scales larger 
than the typical virial radii of halos \citep{cooray_sheth_02}.  
This is equivalent to assuming that the halo density profile 
is a delta function when computing the two halo term 
\footnote{We found that this approximation is sufficient to 
study the clustering of LBGs}. 
In this case, using Eq.~(\ref{eq:xi_hh_full}), the 2-halo term for galaxy 
correlation function can be written as 
\bea
1+\xi^{\mathrm 2h}_{\mathrm g}(r,m, z) =  \df{1}{n_{\mathrm g}(m,z)} \int dM' \df{dn}{dM} (M') N_{\text{g}}(M',m,z) \nonumber \\ 
      \int dM'' \df{dn}{dM}(M'') N_{\text{g}}(M'',m,z) \left[1+\xi_{\mathrm{hh}}(r|M',M'', z) \right],  
\label{eqn:xi-2h}
\eea

\noindent where 
\be
n_{\mathrm g}(m,z) = \int dM \df{dn}{dM}(M,z) N_{\text{g}}(M,m,z) 
\ee
is the number density of LBGs brighter than apparent magnitude $m$ 
given by Eq.\ref{eq:totgal}. 
We use the fitting function of \cite{tinker_kravtsov_08} for the 
halo mass function, $dn/dM(M,z)$, 
as this is found to be in excellent agreement with the halo mass 
function measured from N-body simulations, which is used to calibrate the 
non-linear halo bias.
We have also incorporated the halo exclusion in Eq.~(\ref{eqn:xi-2h}) through the $\Theta$ 
function in  Eq.~(\ref{eq:xi_hh_full}). 

\subsection{The 1-halo term}
The 1-halo term is computed as in J13, 
assuming that the radial distribution of satellite galaxies 
inside a parent halo follows the dark matter density 
distribution \citep{cooray_sheth_02}. 
We use the Navarro-Frenk-White (NFW) profile for the 
density distribution in a dark matter halo \citep{NFW_97}. 
The halo concentration parameter for computing NFW fitting functions is 
taken from \cite{klypin_2011_concp}. 
In this case the 1-halo term is given by (\citealt{sheth_hui_01,tinker_weinberg_05})
\bea
\xi^{\mathrm 1h}_g(r,m,z) = \f{1}{n_g^2}\int dM~ \df{dn}{dM}(M,z) \times  \Big[ 2 N_{\text{cen}}(M,m,z) \nonumber \\ 
    N_{\text{s}}(M,m,z)   \f{\rho_{\text{NFW}}(r,M,z)}{M} + \nonumber \\
     {N_{\text{s}}}^2(M,m,z) \rangle \f{\lambda_{\text{NFW}}(r,M,z)}{M^2} \Big].
\label{eqn:xir-1h}
\eea
Here $\rho_{\text{NFW}}$ is the NFW profile of dark matter density inside a 
collapsed halo and $\lambda_{\text{NFW}}$ is the convolution of this density 
profile with itself \citep{sheth_hui_01}.  

\begin{table*}
\label{table1}
\begin{tabular}{|c c c c| cc|cc|cc|}\hline
\mcn{4}{|c|}{}   &\mcn{4}{c|}{ $\xi_{\mathrm g}$ in the range $0.1\leq \text{r/ $h^{-1}$ Mpc}\leq 50$} \B \B \B  &\mcn{2}{c|}{} \T\T \\
\cline{5-8}
\mcn{4}{|c|}{}  &\mcn{2}{c|}{$r_0$ (Mpc/$h$)} &\mcn{2}{c|}{$\gamma$}  &\mcn{2}{c|}{$b_{\mathrm g}= \sqrt{\df{\xi_{\mathrm g}}{\xi_{\mathrm{mm}}}}$}  \B \B \B   \T\T \\
\cline{1-10}
$z$ &$m$  &$M_{AB}$ &$M_{\text{av}}/M_\odot$  &linear &non-linear &linear &non-linear  &linear &non-linear \T \B\B \B\B\\
\hline  
\T~  &25.5 &-21.0  &$7.0\times10^{11}$   &4.41 &7.43   &1.55 &2.06   &5.85 &6.53 \\
  5  &26.0 &-20.5  &$4.5\times10^{11}$   &3.79 &6.36   &1.53 &1.97   &5.31 &5.80\\
\B~  &26.5 &-20.0  &$2.9\times10^{11}$   &3.25 &5.39   &1.51 &1.89   &4.82 &5.18\\

\T~  &25.0 &-21.1  &$9.9\times10^{11}$   &4.37 &6.71   &1.58 &1.96   &4.72 &5.16\\
  4  &25.5 &-20.6  &$6.1\times10^{11}$   &3.79 &5.72   &1.56 &1.89   &4.29 &4.61\\
\B~  &26.0 &-20.1  &$3.9\times10^{11}$   &3.29 &4.85   &1.54 &1.82   &3.90 &4.14\\

\T~  &24.5 &-21.1  &$1.5\times10^{12}$   &4.26 &6.10   &1.60 &1.93   &3.61 &3.84\\
  3  &25.0 &-20.6  &$9.0\times10^{11}$   &3.67 &5.13   &1.59 &1.86   &3.23 &4.40\\
\B~  &25.5 &-20.1  &$5.5\times10^{11}$   &3.20 &4.36   &1.58 &1.81   &2.92 &3.05\\
\hline
\end{tabular}
\caption{Column 1: redshift of the galaxy sample; 
Columns 2 and 3: apparent and absolute magnitude limits of the galaxy sample.
Column 4: mean halo mass of galaxies in the sample as given in J13. 
Columns 5 and 6: the correlation length obtained from the best fit power-law 
for the linear and non-linear halo bias models.
Columns 7 and 8: the predicted power-law index of the galaxy correlation function in the range 
$0.1\leq \text{r/Mpc}\leq 50$ for the linear and the non-linear halo bias models.
Columns 9 and 10: the galaxy bias at 8 $h^{-1}$ Mpc, predicted by the 
linear and the non-linear halo bias models. 
}
\label{tab1}
\end{table*}

\subsection{The total spatial correlation function}
The galaxy correlation function is obtained by adding the 
1-halo and the 2-halo terms 
\be 
 \xi_{\mathrm g}(r, m,z) =\xi^{\mathrm 1h}(r,m,z) + \xi^{\mathrm 2h}(r,m,z).
\ee

In Fig.~\ref{fig:xir} we show the predicted spatial correlation functions of LBGs 
at $z=3,4$ and $5$.
Each panel corresponds to a given redshift and threshold apparent magnitude. 
Fig.~\ref{fig:xir} clearly shows that, at each redshift, the linear halo bias 
model prediction for $\xi_{\mathrm g}$ breaks away from a power-law on scales from 
$0.1$ to $10 h^{-1}$ Mpc. 
On similar scales, the non-linear halo bias model predicts much stronger 
clustering compared to the linear halo bias model.  
As a result, the deviation of $\xi_{\mathrm g}$ from a power-law 
with the non-linear halo bias model is far smaller than is the case for  
the linear halo bias model. 
We also find that the discrepancy between the predictions of the two 
models increases with redshift and luminosity. In fact, it is clear from Fig.~\ref{fig:xir} that, at 
$z=5$, the linear halo bias model underpredicts the correlation functions 
by more than an order of magnitude for $0.5 \leq \text{r/($h^{-1}$ Mpc)} \leq 1$ 
compared to the non-linear halo bias model. 
This clearly shows that the non-linear bias of dark matter halos must 
be properly accounted for in order to understand the clustering of high-z galaxy 
samples that are characterized by low number densities.

We have fitted the $\xi_{\mathrm g}(r)$ predicted by the linear and 
non-linear halo bias models with a power-law $\xi_{\mathrm g}(r) = (r/r_0)^{-\gamma}$ 
where $r_0$ is the 
correlation length and $\gamma$ is the power-law index. 
The best fitting $r_0$ and $\gamma$, obtained by fitting 
$\xi_{\mathrm g}$ in the range $0.1 \leq \text{r/$h^{-1}$ Mpc} \leq 50$, 
are tabulated in Table \ref{tab1}. 
It is clear from  columns 5 and 6 of Table \ref{tab1} that the correlation lengths 
predicted by the non-linear halo bias model are systematically higher than those obtained 
from the linear halo bias model. This is a direct consequence of the increased clustering 
of LBGs on quasi-linear scales in the non-linear halo bias model. 
In particular, 
for the brightest LBG sample at $z=5$, the predicted correlation length in 
the non-linear halo bias model is larger by a factor $1.7$ compared to 
the linear bias prediction.
Furthermore, $r_0$ increases with redshift and galaxy luminosity. 
For example, the correlation length at $z=5$ is larger by a factor $\sim 1.2$ compared to that at $z=3$ for a given apparent limiting magnitude. This is consistent with the 
observational results of \citet{hildebrandt_09_acf} who found a similar 
evolution of $r_0$ with redshift and limiting magnitude. 

The power-law slopes of the predicted $\xi_{\mathrm g}$ for the linear and the non-linear 
bias models are given in columns 7 and 8 of Table \ref{tab1}. 
It is clear from Table \ref{tab1} that, the linear halo bias model predicts 
$\gamma = 1.5-1.6$. On the other hand $\gamma$ predicted by non-linear and 
scale-dependent bias model ranges typically from $1.8-2.1$. This is remarkably 
consistent with the values of the power-law index ($\gamma \sim 1.7- 2.1$) 
inferred from high redshift observations \citep{ouchi_hamana_05_acf,
lee_giavalisco_06_acf, coil_06, hildebrandt_09_acf, 
savoy_11_acf, bian_2013_acf, barone_2014_acf, durkalec_2015_acf, 
harikane_2016_acf}. 

Finally, we compare the galaxy bias predicted on large scales by the linear 
and non-linear models. 
We first note that the large scale LBG bias has been measured by numerous 
authors at different scales using alternative definitions. 
Several set the scale at which the bias of LBGs is measured 
to be 8 $h^{-1}$ Mpc \citep{ouchi_04_acf,adelberger_05,yoshida_08,hildebrandt_09_acf}. 
Following these, we have defined the galaxy bias, $b_{\mathrm g}$ at a scale of 
8 $h^{-1}$ Mpc as
\be
{b_{\mathrm g}} = \sqrt{\df{\xi_{\mathrm g}(r)}{\xi_{\mathrm{mm}}(r)}}\Biggr|_{r=8 h^{-1}\text{Mpc}}
\label{eq:gbias}
\ee
The galaxy bias estimated using Eq.~\ref{eq:gbias} for the linear and non-linear 
bias models are tabulated in columns 9 and 10 of Table~\ref{tab1} respectively. 
Interestingly, we find that $b_{\mathrm g}$ predicted by the 
non-linear halo bias model are consistently higher than the linear halo bias predictions. 
Specifically, in the redshift range $3-5$, the galaxy bias is larger 
by $4-12$\% in the non-linear halo clustering model compared to the 
linear halo bias model. 
It is remarkable that the non-linear dark matter halo 
clustering results in such an increment in the galaxy bias 
even on a scale of 8 $h^{-1}$ Mpc at high-z where matter fluctuations 
are still in the linear regime\footnote{For example $\sigma_8 = 0.175$ at $z=5$.}.

\subsection{The angular correlation function}

Now we compute the angular correlation function, a direct measurable 
of galaxy clustering, from the spatial correlation function. Several 
authors have measured the angular correlation function of LBGs brighter 
than a given UV luminosity in the 
redshift range $3 \leq z \leq 5$ \citep{ouchi_hamana_05_acf,
lee_giavalisco_06_acf, kashikawa_06_acf,coil_06, hildebrandt_09_acf, 
wake_11,savoy_11_acf, bian_2013_acf, barone_2014_acf, 
durkalec_2015_acf, harikane_2016_acf}. Among these studies 
\cite{hildebrandt_09_acf} uses the largest LBG sample in the 
redshift range $3\leq z \leq 5$. This allows them to estimate 
the angular correlation functions with small error bars for a 
range of magnitudes. Hence we will use the \cite{hildebrandt_09_acf} 
data when comparing our model predictions with observations.

The luminosity dependent angular correlation function 
$w(\theta, m,z)$ is computed from the spatial correlation function using Limber's  equation \citep{peebles_80}
\be
w(\theta, m,z) = \int_0^\infty dz'~ N(z') \int_0^{\infty} dz''~N(z'') 
\xi_{\mathrm g}\left(z,r(\theta; z',z'')\right) 
\label{eqn:limber}
\ee
where $r(\theta; z',z'')$ is the comoving separation between two points 
at $z'$ and $z''$ subtending an angle $\theta$ with respect to an observer. 
To compute the angular correlation functions we used the normalized redshift 
selection function, $N(z)$, from ~\citet{hildebrandt_09_acf} ($BC_{\rm sim}$ 
redshift distribution; see Table 4 and Figure 5 of their paper). 
Furthermore, in Eq.~(\ref{eqn:limber}), the spatial two-point correlation 
function $\xi_{\mathrm g}(r,z)$ is always evaluated at the central redshift of the 
selection function $N(z)$. 

In Fig.~\ref{fig:acf}, we show in solid black lines, the angular correlation 
functions of high-z LBGs computed after incorporating the non-linear halo bias. 
The results are presented for $3 \leq z \leq 5$ and for a wide range of 
magnitude limits. Fig.~\ref{fig:acf} also shows the angular  
correlation function of LBGs computed using the linear halo bias model  
given in J13 as dashed red lines.  

First, we note that the clustering predictions using linear halo bias compare 
reasonably well with the observed data for $\theta > 100''$. 
However, on intermediate 
angular scales ($5'' \leq \theta \leq 100''$), the linear halo bias predictions do  
not match the observed clustering.    
Here, there is a clear break in the predicted angular correlation functions 
around $\theta \sim 10''$, due to which the correlation 
functions look like a double power-law. 
This scale corresponds to the transition from the 2-halo to 1-halo contributions  
to the clustering. Furthermore, the discrepancy between theory and observations 
seems to increase for brighter and higher-z galaxy samples.

The strong feature in the predicted LBG clustering at intermediate 
angular scales can also be found in earlier studies using the halo 
model \citep{hamana_04,ouchi_hamana_05_acf, lee_09,charles_13_clustering,
bian_2013_acf, harikane_2016_acf}. These studies use the linear halo bias model
for the 2-halo contribution  
which in turn results in a very strong deviation of the predicted angular 
correlation function from a power-law.
It is possible to increase the clustering strength over $5'' \leq \theta \leq 100''$ 
by changing the HOD given by $N_{\text{g}}$ in Eq.~(\ref{eqn:xi-2h}). However, this will 
simply rescale $\xi_{\mathrm g}$ upwards and as a result, on large scales, 
the linear halo bias 
model will overpredict the observed clustering. 

Our model predictions using non-linear halo bias significantly 
increase the clustering strength of LBGs only in the angular range  
$5'' \leq \theta \leq 100''$, on the whole greatly improving the 
agreement with observational measurements. 
It is also clear that the boost in the clustering 
strength due to the non-linear halo bias on this scale is larger for higher redshift 
and for brighter galaxy samples. 
As a result, the feature predicted by the linear halo bias model at the 2-halo 
to 1-halo transition region (at $\theta \sim 10''$),  
weakens and the correlation function is closer to a power-law. 
Therefore, we conclude that the non-linear bias of high-z dark matter halos 
plays a major role in reshaping the correlation functions of high-z galaxies, 
thereby providing better agreement with the observational data.

We also note from Fig.~\ref{fig:acf} that, for fainter galaxy 
samples ($m = 25.0$ and $25.5$) at $z=3$, the non-linear bias model 
slightly overpredicts the clustering on intermediate angular scales, and 
the linear bias model provides a better fit to the data. 
One could in principle investigate whether this is due to uncertainties 
in the HODs of those samples by using alternative HODs derived from 
more sophisticated galaxy formation models, but such an analysis is 
beyond the scope of this paper. 
However, apart from these two cases, the clustering measurements 
for all of the other galaxy samples are much better fit by 
the non-linear halo bias model than the linear halo bias model.

Finally, it is clear from Fig.~\ref{fig:acf} that, on scales smaller than $\sim 10''$ 
and larger than $\sim 100''$, the correlation functions of LBGs, as 
predicted by linear and non-linear halo bias models,  agree well with each other. 
As noted earlier, for $\theta \leq  10''$, the non-linear halo bias does not 
affect the correlation function because the clustering on these scales is 
dominated by the 1-halo contribution due to pairs of galaxies sitting 
inside the same dark matter halo. 
On the other hand, on large scales $\zeta(r,M,z) \xrightarrow{~} 1$ 
and hence the expression for the non-linear halo bias in Eq.~(\ref{eq:nlbias}) reduces back 
to the linear halo bias. 
As a result, for $ \theta \geq  100''$, the effect of scale-dependent non-linear 
bias is insignificant for LBG clustering.

\section{Discussion and conclusions}

We have investigated the non-linear clustering of high-z LBGs 
in the redshift range $3 \leq z \leq 5$ using the halo model of large 
scale structure. 
Specifically, we find that incorporating the non-linear halo bias of these halos 
in the halo model is of utmost importance for understanding the clustering 
of LBGs in the quasi-linear regime corresponding to angular 
scales of $5'' \leq \theta \leq 100''$. 

Our work is motivated by \citet{charles_13_clustering} who showed that 
the halo model predictions using linear halo bias for halo clustering fail to explain the 
observed clustering amplitude of LBGs at $3 \leq z \leq 5$ on 
angular scales $5'' \leq \theta \leq 100''$. 
The linear halo bias models underpredict LBG clustering on these 
scales by an order of magnitude.   
As a result the predicted LBG correlation functions in the 
above redshift range depart significantly from a 
power-law in contrast with observations. 

We address this issue using the model of \cite{jose_16} who 
investigated the non-linear clustering of 
high-z dark matter halos in the redshift range and on the scales of 
interest. 
In particular, these authors provide an analytic fitting function for 
the scale-dependent non-linear halo bias of high-z halos on quasi-linear 
scales, which is a function of four quantities that can be readily  
computed for any cosmology.  
Using this, we find that, at $z=4$, the non-linear bias of dark matter 
halos of mass $\sim 10^{12} M_\odot$, that host typical LBGs at this  
redshift, is quite significant on quasi-linear scales. 
As a result, the clustering amplitude is enhanced by up to a factor of 20 at a 
scale of $0.5$ $h^{-1}$Mpc. 

We combined the analytic fitting formula for non-linear halo bias given by \cite{jose_16} 
with the halo model to predict the spatial 
correlation function of LBGs. 
For this we used the HOD of LBGs given by \citet{charles_13_clustering}, 
computed using their galaxy formation model.  
The corresponding predicted correlation function shows much stronger clustering 
on scales $0.1 - 10$ $h^{-1}$ Mpc compared to the predictions of the 
linear halo bias model. 
Furthermore, the difference between the models increases with redshift and 
galaxy luminosity. 
The resulting galaxy correlation function in the non-linear halo bias model 
are much closer to a power-law than the predictions of the 
linear halo bias model. 
The corresponding correlation lengths of LBGs are consistently larger in 
the non-linear halo bias model compared to the linear halo bias model. 
For example, at $z=5$ and for a limiting magnitude of $m=25.5$, the predicted 
correlation length in the non-linear halo bias model is larger than 
the linear halo bias model by $70$\%.     
Moreover, the non-linear halo bias model predicts a power-law index of $\gamma 
\sim 1.8 -2.1$ compared to $\gamma \sim 1.5-1.6$ obtained from the linear halo bias model. 
Remarkably, the power-law index estimated using the non-linear halo bias 
model at high redshifts compares very well with 
the values deduced from several high redshift observations. 

The spatial correlation functions are then used to compute the angular 
correlation functions of LBGs. We find that the non-linear bias of dark 
matter halos significantly boosts the clustering 
of LBGs on angular scales of interest ($10'' \leq \theta \leq 100''$). 
The resulting LBG correlation functions provide a much better 
fit to the observed data on these angular scales compared to the predictions of 
the linear halo bias model, except for the fainter galaxy samples 
at $z=3$. 
The effect of non-linear halo clustering is also found to increase with 
redshift and galaxy luminosity. 
This is expected because the non-linear bias is larger for 
more massive and higher redshift dark matter halos. 
This, in turn, reshapes the angular correlation functions of LBGs in the 
redshift range $3$ to $5$ into an approximate power-law over the 
entire angular scale. 
While at low $z$, a power-law $\xi_{\mathrm g}(r)$ is achieved by the fine-tuning of 
several ingredients including the HOD and the 
dark matter clustering \citep{watson_berlind_11}, we believe at high-$z$ the 
non-linear halo bias plays a critical role in making the shape of 
the galaxy correlation function close to a power-law. 

Finally, the predicted LBG bias at 8 $h^{-1}$ Mpc is larger 
in the non-linear halo bias model compared to the linear halo bias predictions 
for all redshifts and limiting magnitudes.
In particular, for $3\leq z \leq 5$, we find a $4-12$\% increase in the 
galaxy bias on this scale in the non-linear halo clustering models.
This is because the effective bias on this scale is not 
the linear asymptotic bias, but is still subject to a substantial 
scale dependent effect, as shown by Fig.~\ref{fig:bnl}. 
Since several studies measure the galaxy bias of LBGs at 8 $h^{-1}$ Mpc 
\citep{ouchi_04_acf,adelberger_05,yoshida_08,hildebrandt_09_acf}, 
it is very important to include the non-linear halo bias in the halo 
model of LBG clustering for a reliable comparison of data and theory. 

The observed clustering of LBGs has been used in numerous studies 
as a probe of the physics of high-redshift galaxy formation 
\citep{hamana_04,ouchi_hamana_05_acf, lee_09, 
ouchi_lae_10,charles_13_lae_clustering, cooke_2013_acf, bian_2013_acf, 
durkalec_2015_acf, harikane_2016_acf}. 
These studies used the linear halo bias model to place constraints on several 
interesting quantities such as the average mass of halos 
hosting LBGs, the duty cycle of star formation activity and the 
fraction of satellite galaxies at high redshifts. 
The use of the linear halo bias model potentially introduces  
systematics in the inferred values of these quantities. 
For example, for LBGs brighter than $m=26.5$ at $z=5$, focusing on just 
the measured galaxy clustering at $\theta \sim 14''$ rather than the 
whole correlation function (see Fig.~\ref{fig:acf}), comparing to the 
clustering expected in the dark matter suggests a galaxy bias of 
$b_{\mathrm g} \sim 9.3$. In the linear model, this bias translates 
into a mean halo mass of $\sim 4.2 \times 10^{12}$ M$_\odot$.  
However, if instead we use the non-linear halo bias model to interpret the 
clustering at this fixed angular scale, part of the difference in 
amplitude between the galaxy clustering and the dark matter clustering 
is due to the non-linear effects. This means that the asymptotic bias on 
large scales is smaller than in the linear halo bias model, implying a lower mean halo 
mass $\sim 3.9 \times 10^{11}$ M$_\odot$, which is an order of magnitude 
smaller than that inferred from the linear halo bias model. 
Therefore, it is very important to incorporate the 
non-linear halo bias of high-z halos in the clustering models of LBGs 
to better explore the physics of galaxy formation 
and cosmology. 

Even though we have focussed on the importance of the non-linear halo bias 
for the clustering of LBGs at $3 \leq z \leq 5$, one can expect similar 
trends in the clustering of rare, high-redshift galaxies for other 
sample selections. In particular, strongly non-linear clustering is 
expected for high-z quasars, Lyman-$\alpha$ emitters, dusty star-forming 
galaxies and redshifted 21 cm signals from the pre-reionization era. 
The extended version of the halo model we have introduced that incorporates 
the non-linear bias of dark matter halos will be a useful tool for the robust 
interpretation of such measurements.


\section*{Acknowledgments}
CJ acknowledges a Durham COFUND Junior Research Fellowship (4958) supported 
by the DIFeREns2 project which has received funding from the European Union's 
Seventh Framework Programme for research, technological development and 
demonstration under grant agreement no 609412. 
This work was supported by the Science and Technology Facilities Council 
[ST/L00075X/1]. This work used the DiRAC Data Centric System at Durham 
University, operated by the ICC on behalf of the STFC DiRAC HPC 
Facility (www.dirac.ac.uk). This equipment was funded 
by BIS National E-infrastructure capital grant ST/K00042X/1, STFC capital 
grant ST/H008519/1, and STFC DiRAC Operations grant ST/K003267/1 and 
Durham University. DiRAC is part of the UK's National E-Infrastructure. 

\def\aj{AJ}%
\def\actaa{Acta Astron.}%
\def\araa{ARA\&A}%
\def\apj{ApJ}%
\def\apjl{ApJ}%
\def\apjs{ApJS}%
\def\ao{Appl.~Opt.}%
\def\apss{Ap\&SS}%
\def\aap{A\&A}%
\def\aapr{A\&A~Rev.}%
\def\aaps{A\&AS}%
\def\azh{AZh}%
\def\baas{BAAS}%
\def\bac{Bull. astr. Inst. Czechosl.}%
\def\caa{Chinese Astron. Astrophys.}%
\def\cjaa{Chinese J. Astron. Astrophys.}%
\def\icarus{Icarus}%
\def\jcap{J. Cosmology Astropart. Phys.}%
\def\jrasc{JRASC}%
\def\mnras{MNRAS}%
\def\memras{MmRAS}%
\def\na{New A}%
\def\nar{New A Rev.}%
\def\pasa{PASA}%
\def\pra{Phys.~Rev.~A}%
\def\prb{Phys.~Rev.~B}%
\def\prc{Phys.~Rev.~C}%
\def\prd{Phys.~Rev.~D}%
\def\pre{Phys.~Rev.~E}%
\def\prl{Phys.~Rev.~Lett.}%
\def\pasp{PASP}%
\def\pasj{PASJ}%
\def\qjras{QJRAS}
\def\rmxaa{Rev. Mexicana Astron. Astrofis.}%
\def\skytel{S\&T}%
\def\solphys{Sol.~Phys.}%
\def\sovast{Soviet~Ast.}%
\def\ssr{Space~Sci.~Rev.}%
\def\zap{ZAp}%
\def\nat{Nature}%
\def\iaucirc{IAU~Circ.}%
\def\aplett{Astrophys.~Lett.}%
\def\apspr{Astrophys.~Space~Phys.~Res.}%
\def\bain{Bull.~Astron.~Inst.~Netherlands}%
\def\fcp{Fund.~Cosmic~Phys.}%
\def\gca{Geochim.~Cosmochim.~Acta}%
\def\grl{Geophys.~Res.~Lett.}%
\def\jcp{J.~Chem.~Phys.}%
\def\jgr{J.~Geophys.~Res.}%
\def\jqsrt{J.~Quant.~Spec.~Radiat.~Transf.}%
\def\memsai{Mem.~Soc.~Astron.~Italiana}%
\def\nphysa{Nucl.~Phys.~A}%
\def\physrep{Phys.~Rep.}%
\def\physscr{Phys.~Scr}%
\def\planss{Planet.~Space~Sci.}%
\def\procspie{Proc.~SPIED}%
\let\astap=\aap
\let\apjlett=\apjl
\let\apjsupp=\apjs
\let\applopt=\ao

\bibliographystyle{mn2e}	
\bibliography{ref.bib,sfr.bib}		

\end{document}